\documentclass[apj]{emulateapj}

\shorttitle{Resolving loops in the corona}
\shortauthors{Brooks et al.}

\begin{document}

\title{Solar Coronal Loops Resolved by Hinode and SDO }             
\author{David H. Brooks \altaffilmark{1}} 
\affil{College of Science, George Mason University, 4400 University Drive, Fairfax, VA 22030}                                  
\email{dhbrooks@ssd5.nrl.navy.mil}
\author{Harry P. Warren}
\affil{Space Science Division, Naval Research Laboratory, Washington, DC 20375}
\author{Ignacio Ugarte-Urra}
\affil{College of Science, George Mason University, 4400 University Drive, Fairfax, VA 22030}                                  
\altaffiltext{1}{Present address: Hinode Team, ISAS/JAXA, 3-1-1 Yoshinodai, Chuo-ku, Sagamihara, Kanagawa 252-5210, Japan}

\begin{abstract}

Despite decades of studying the Sun, the coronal heating problem remains unsolved. One fundamental issue is that we do not know the spatial scale of the coronal heating mechanism. At a spatial resolution of 1000 km or more it is likely that most observations represent superpositions of multiple unresolved structures. In this letter, we use a combination of spectroscopic data from the {\it Hinode} EUV Imaging Spectrometer (EIS) and high resolution images from the Atmospheric Imaging Assembly (AIA) on the {\it Solar Dynamics Observatory} to determine the spatial scales of coronal loops. We use density measurements to construct multi-thread models of the observed loops and confirm these models using the higher spatial resolution imaging data. The results allow us to set constraints on the number of threads needed to reproduce a particular loop structure. We demonstrate that in several cases million degree loops are revealed to be single monolithic structures that are fully spatially resolved by current instruments. The majority of loops, however, must be composed of a number of finer, unresolved threads; but the models suggest that even for these loops the number of threads could be small, implying that they are also close to being resolved. These results challenge heating models of loops based on the reconnection of braided magnetic fields in the corona. 
\end{abstract}

\section{Introduction}

While it is almost universally accepted that turbulent motions in the photosphere associated with magnetoconvection provide the energy that heats the corona, there is no consensus as to whether this energy is transferred by waves \citep[e.g.][]{vanballegooijen_etal2011,mcintosh_etal2011}, the dissipation of magnetic stresses through magnetic reconnection \citep{parker_1983}, or injected directly by chromospheric jets  \citep{depontieu_etal2009,depontieu_etal2011}. Determining the details of the coronal heating mechanism would be greatly simplified if we could resolve individual coronal loops. The spatial resolution of instruments used to observe the corona, however, has generally been 1000 km or more. The solar photosphere, in contrast, shows structure on scales of 100 km or less. We therefore cannot be sure that the properties we measure in the corona actually correspond to those of individual loops. Arguably the most important question we need to answer is this: at what spatial scale do we resolve coronal loops? 

Previous work on this topic has shown that the temperature distribution in coronal loops is often narrow, suggesting that loops could be resolved \citep{aschwanden&nightingale_2005,warren_etal2008a}. These studies, however, either assumed that there is no substructure \citep{aschwanden&nightingale_2005} or simply computed a filling factor \citep{warren_etal2008a}. The latter requires assumptions about the complex relationship between loop radius and measured width to crudely approximate the geometry. Here we use density measurements to infer the spatial scale of the emission and construct multi-thread models that can reproduce the cross-field intensity profiles observed with both EIS and higher spatial resolution AIA image data. With the geometry determined, we report the actual sizes of the strands. We generally find that a single thread implies loops that are much narrower than we observe, indicating that most loops are unresolved by current instrumentation. Our models show, however, that the observations can be reproduced with a relatively small number of threads several hundred km in width, and we do find a few cases of individual loops that are fully spatially resolved. This suggests that the ability to resolve coronal loops routinely will be achieved by the next generation of coronal instruments. 

\begin{figure}[t!]
\centerline{%
\includegraphics[viewport=150 130 440 750,clip,scale=0.85]{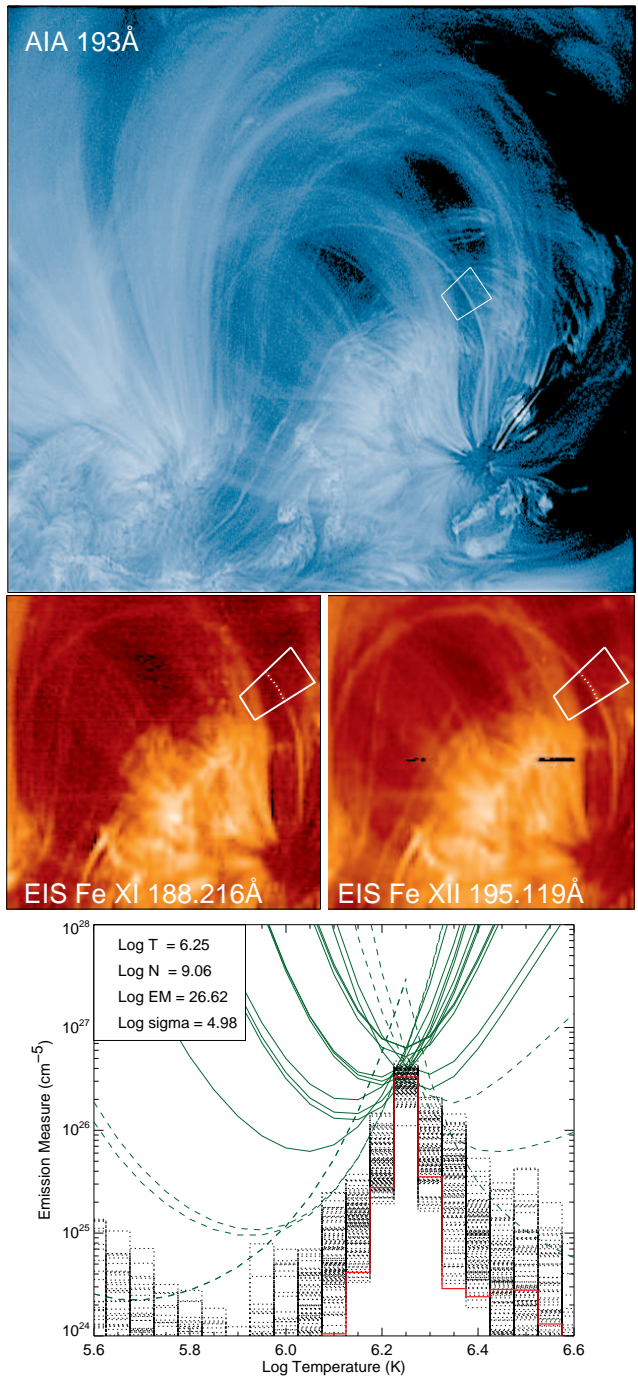}}
\caption{AIA 193\,\AA\, filter image (upper panel) and EIS slit raster scans in
\ion{Fe}{11} 188.216\,\AA\, and \ion{Fe}{12} 195.119\,\AA\, (center panels) showing a 
fully spatially resolved loop. This is 
loop $\#$19 in Table \ref{table}. The traced
loop segments are indicated.
The best-fit EM distribution computed by the MCMC algorithm is shown in red in the lower panel. Random realizations of the solution are shown as dotted lines. The green curves are EM loci (observed intensity normalized by the $G(T,n)$ function) for each line. Those lines that are not well correlated with \ion{Fe}{12} 195.119\,\AA\, are indicated with dashed curves. The peak temperature, emission measure, electron density, and Gaussian width of the EM are given in the legend.
}
\label{fig1}
\end{figure}
\section{Data Reduction and Analysis Methods}
Understanding the observed intensity for optically thin coronal emission requires information about the physics of the atomic transitions being observed and the geometry of the emitting structure. If we represent an observed loop as the superposition of a series of cylindrical flux tubes, the loop intensity is
\begin{equation}
I = G(T,n) n^2 V/l^2 = G(T,n) n^2 N \pi r^2/l
\label{equation1}
\end{equation}
where  $I$ is the observed intensity across a loop, $G(T,n)$ is the normalized response of the plasma as a function of the electron temperature ($T$) and density ($n$), and $V$ is the volume of the emitting structure within an instrumental resolution element of area $l^2$. The second expression assumes that the emitting structures are a series of $N$ identical loops of radius $r$. This equation allows us to derive information on the cross-field spatial scale of coronal loops, and therefore to directly address our question.

EIS on {\it Hinode} \citep[][]{kosugi_etal2007,culhane_etal2007a} is uniquely capable of determining the spectroscopic diagnostic quantities required to perform this analysis. EIS routinely observes as many as 40 strong coronal emission lines in the wavelength ranges from 166--212\,\AA\, and 246--292\,\AA\, at approximately 22 m\,\AA\, spectral resolution. Several of these emission lines form density sensitive pairs which can be used to determine the electron density. They also cover a wide range of temperatures and their intensities allow us to compute the distribution of temperatures in a loop: the emission measure (EM) distribution. This methodology is well known and has already been applied to EIS observations of coronal loops by previous authors \citep{warren_etal2008a,tripathi_etal2009,schmelz_etal2010,delzanna_etal2011}. 

AIA on the Solar Dynamics Observatory \citep[SDO,][]{pesnell_etal2012,lemen_etal2011} observes the solar corona in 6 relatively broad wavelength ranges (1--11\,\AA\, bandpasses). AIA has very high spatial resolution, 824 km full width half maximum (FWHM) point spread function (PSF) compared with 1800 km for EIS \citep{grigis_etal2012,lang_etal2006}, but fewer plasma diagnostics. 

Here we collect a sample of 20 relatively isolated active region loops and use the AIA data to confirm the results we obtain with EIS.
The EIS data have been processed for instrumental effects and converted to physical units (erg/cm2/s/steradian) using the SolarSoftware (SSW) routine eis\_prep. 
A specially designed observing sequence is used. This sequence uses the 1$''$ slit to scan a field-of-view (FOV) of 240$''$ by 512$''$ in coarse 2$''$ steps. The exposure time at each position is 60s. The duration of the scan is therefore about 2 hours and many diagnostic lines are included. Some of these are discussed below. The data are finally re-sampled to an arcsecond scale.

\begin{deluxetable*}{rrrrrrrrrrrrrrrrrr}
\tabletypesize{\scriptsize}
\tablehead{
\multicolumn{1}{c}{\#} &
\multicolumn{1}{c}{Date/Time} &
\multicolumn{1}{c}{$\sigma_w$} &
\multicolumn{1}{c}{$n$} &
\multicolumn{1}{c}{$T$} &
\multicolumn{1}{c}{$EM$} &
\multicolumn{1}{c}{$r_1$} &
\multicolumn{1}{c}{$N$} &
\multicolumn{1}{c}{$r$} &
\multicolumn{1}{c}{\#} &
\multicolumn{1}{c}{Date/Time} &
\multicolumn{1}{c}{$\sigma_w$} &
\multicolumn{1}{c}{$n$} &
\multicolumn{1}{c}{$T$} &
\multicolumn{1}{c}{$EM$} &
\multicolumn{1}{c}{$r_1$} &
\multicolumn{1}{c}{$N$} &
\multicolumn{1}{c}{$r$} 
}
\tablewidth{0pt}
\tablecaption{Observed and Simulated Properties of Active Region Loops Observed with EIS}
\startdata
   1 & 01/21 13:57 & 1103 & 9.68 & 6.25 & 27.6 & 550 & 6 & 225 & 11 & 04/29 01:23 &  1294 & 9.12 & 6.25 & 27.3 & 1450 & 1 & 1450 \\
   2 & 01/30 20:11 & 859 & 9.40 & 6.30 & 27.0 & 690 & 2 & 404 & 12 & 05/06 13:55 &  1208 & 9.49 & 6.20 & 27.3 & 650 & 4 & 330 \\
   3 & 03/18 09:34 & 1468 & 9.22 & 6.20 & 27.1 & 935 & 8 & 335 & 13 & 06/14 00:52 &  1222 & 9.21 & 6.25 & 27.2 & 1095 & 5 & 490 \\
   4 & 03/18 11:05 & 1899 & 9.13 & 6.15 & 27.5 & 1980 & 6 & 815 & 14 & 07/02 03:32 &  1388 & 9.46 & 6.20 & 27.2 & 620 & 5 & 280 \\
   5 & 04/15 00:41 & 1526 & 9.45 & 6.20 & 27.5 & 850 & 7 & 325 & 15 & 07/02 03:53 &  1186 & 9.02 & 6.20 & 27.1 & 1510 & 1 & 1510 \\
   6 & 04/15 01:41 & 682 & 9.46 & 6.20 & 26.7 & 300 & 1 & 300 & 16 & 09/02 13:25 &  1711 & 9.41 & 6.20 & 27.7 & 1040 & 5 & 470 \\
   7 & 04/21 12:02 & 1426 & 9.06 & 6.30 & 27.2 & 1730 & 3 & 990 & 17 & 09/02 13:54 &  1539 & 9.28 & 6.25 & 26.9 & 725 & 6 & 305 \\
   8 & 04/21 12:18 & 1269 & 9.20 & 6.25 & 27.5 & 1255 & 3 & 730 & 18 & 09/17 18:57 &  1859 & 9.67 & 6.25 & 28.3 & 1000 & 5 & 450 \\
   9 & 04/22 08:56 & 1460 & 9.21 & 6.20 & 27.3 & 1220 & 5 & 550 & 19 & 10/14 22:55 &  915 & 9.06 & 6.25 & 26.6 & 830 & 1 & 830 \\
   10 & 04/22 09:00 & 1350 & 9.25 & 6.30 & 27.5 & 1210 & 3 & 700 & 20 & 10/14 23:51 &  990 & 9.58 & 6.25 & 27.3 & 485 & 3 & 280 \\
\enddata
\tablenotetext{}{The dates and times given indicate when EIS was
rastering over the loop segment. All dates are in 2011. The parameter $\sigma_w$ is the Gaussian 
width of the loop cross-field intensity profile in km measured in Fe\,\textsc{xii} 195.119\,\AA. The base-10
logarithm of the electron density ($n$) in cm$^{-3}$, peak temperature ($T$) in Kelvin,
and emission measure ($EM$) in cm$^{-5}$ are given. $r_1$ is the size of
a single thread that matches the observed intensities in km. 
$N$ is the minimum number of simulated threads and $r$ is the maximum thread radius in km. }
\label{table}
\end{deluxetable*}

\begin{figure*}[t!]
\centerline{%
\includegraphics[viewport= 60 100 530 810,clip,scale=0.75,angle=90]{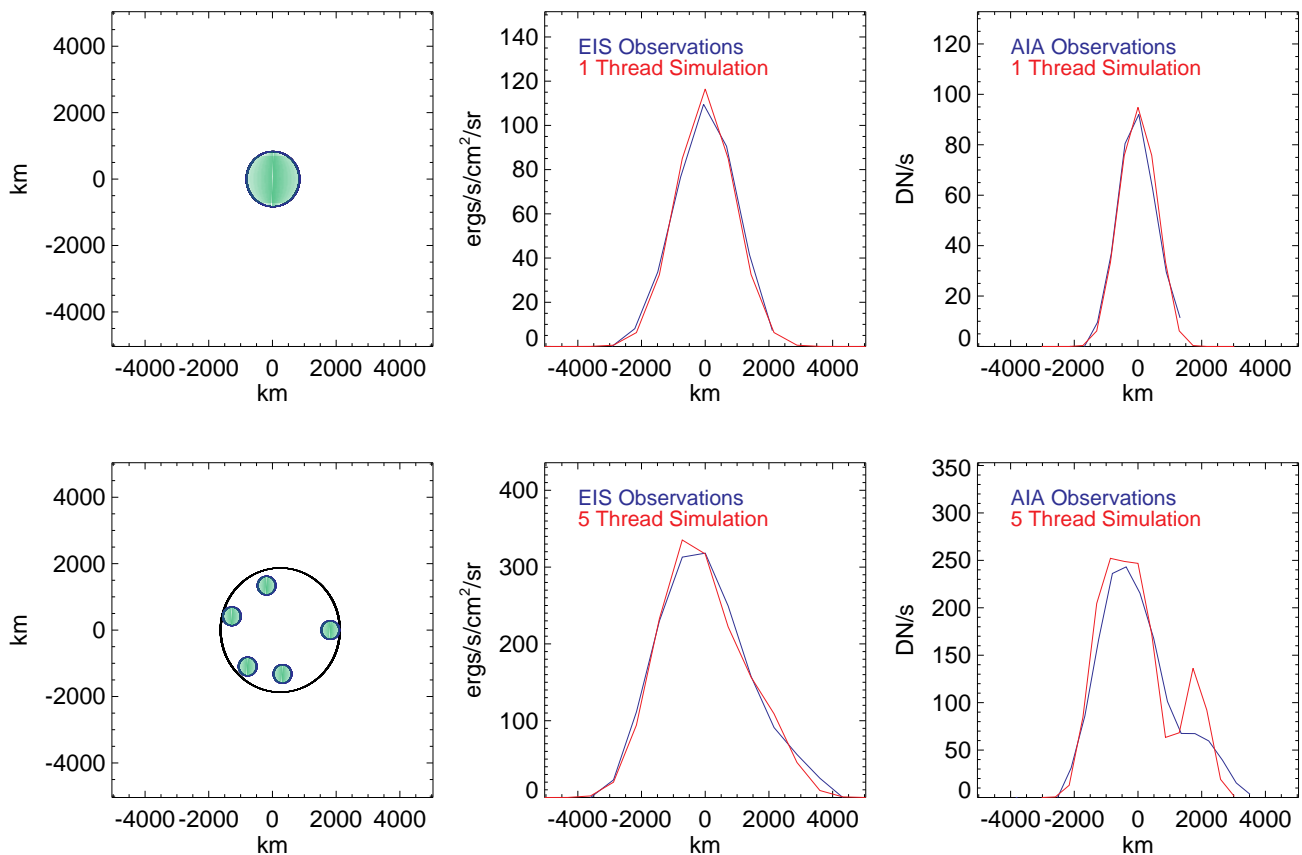}}
\caption{Multi-thread simulations of the spatially resolved loop $\#$19 shown in Figure \ref{fig1} (upper panels)
and the unresolved loop $\#$14 shown in Figure \ref{fig4} (lower panels). The left column shows the simulated
loop envelope (black circle) and the modeled threads (green). The observed cross-loop intensity 
profiles for EIS and AIA are shown in blue, and the simulated profiles are shown in red. 
A single thread completely fills the loop envelope and reproduces the intensity profiles for loop $\#$19.
Five 280 km threads are needed to reproduce the profiles for loop $\#$14.
}
\label{fig2}
\end{figure*}
AIA data were retrieved from the online cut-out service. These data are level 1.5 and have been corrected for flat fields and dark current. Bad pixels and cosmic rays have been cleaned, and the data converted to DN/pixel/s. The data are internally coaligned by mapping to the HMI (Helioseismic and Magnetic Imager) rotation and FOV center. Any residual offset is corrected by additional cross-correlation of images using the SSW routine align\_cube\_correl. The exposure time for each image is 1.7--4s, and we retrieved data covering the EIS FOV for plus/minus 2 hours around the time of the EIS loop observation.

Affine transformation software developed for use with optical observations (which typically require higher accuracy coalignment than possible for coronal images) was modified and used to obtain the relative scaling, rotation, displacement, and shear between the EIS \ion{Fe}{12} 195.119\,\AA\, raster images and the AIA 193\,\AA\, filter image taken at the time the loop is observed. The EIS data are fitted to a double Gaussian function prior to this procedure. The secondary Gaussian component is to take account of the weak blend at 195.18\,\AA. Since the EIS pointing varies with the satellite orbit, the sampling in the slit rasters is not uniform across the FOV. Therefore, any inaccuracy in the AIA-to-EIS image registration is corrected visually as a final step. The EIS and AIA images shown in the Figures in this Letter were treated with a Gaussian sharpening filter with a radius of 3 pixels for presentation, but all analysis was performed on the unfiltered data.

Suitable loops were first identified in the EIS raster data, and relatively isolated portions selected. Once identified, cross-loop intensity profiles perpendicular to the loop axis were automatically extracted and then averaged along the loop within the selected segment following established methodology \citep{warren_etal2008a,aschwanden_etal2008}. We then subtract a co-spatial background component by visually identifying the loop in the intensity profile and fitting a 1st order polynomial between two background pixels. This procedure is applied to the strong \ion{Fe}{12} 195.119\,\AA\, line data, and the same background locations are used for all the other spectral lines. For this analysis, we use spectral lines from consecutive ionization stages of \ion{Fe}{8}--\ion{Fe}{17} that we have found to be reliable in previous EM analysis \citep{warren&brooks_2009,brooks_etal2009}. A Gaussian function is fitted to the background subtracted intensity profile to obtain the loop intensity and FWHM. The electron density is then calculated using the recommended \ion{Fe}{13} 202.044/203.826 ratio, and the EM analysis is made using all the other lines. The CHIANTI database is used for the atomic data including the new ionization fractions \citep{dere_etal1997,dere_etal2009}. Following previous work \citep{warren_etal2008a} we only use those lines, the intensity profiles of which are highly correlated with that of \ion{Fe}{12} 195.119\,\AA. The instrumental calibration uncertainty of 22\% is used as the measurement error of these lines \citep{lang_etal2006}. The other lines have their intensities set to zero, and their uncertainties set to 20\% of the background emission. Finally, for each loop we measured the FWHM of the cross-field intensity profile in both EIS and AIA.

The EM calculation is performed using a Monte Carlo Markov Chain (MCMC) algorithm available in the PINTofALE software package \citep{kashyap&drake_1998,kashyap&drake_2000}. 100 realizations of the solution are calculated. In addition, we separately compute another 100 Monte Carlo simulations to verify that the best-fit emission measure, thermal width, and peak temperature do not change significantly if we perturb the intensities randomly within the calibration uncertainty. As a consistency check, we also fit the data with isothermal and Gaussian emission measure functions. Finally, we ensure that the selected loops satisfy the width criteria suggested by \citet{aschwanden_etal2008}, but we do not adopt their background contrast criteria because we are interested in detecting cases where an EIS loop is composed of several AIA loops. Such cases would be missed if the procedure is optimized to detect single AIA loops. 

The multi-thread models are calculated at a resolution of 1 km using Equation \ref{equation1} assuming an isothermal plasma for each thread and that the threads are of equal radii. The threads are evenly distributed within the loop envelope. The thread intensity profiles are combined to produce a composite profile for the loop, which is convolved with the instrument PSF and resampled to the instrument pixel scale. This profile can be compared to the observations. A broad parameter space of loop envelope and thread radia is explored to find the best match with observed EIS loop intensities and FWHM. Finally, a minimum number of threads compatible with both the EIS and AIA data is determined and their positions slightly perturbed to obtain better agreement with the asymmetric intensity profiles. 

\section{Results and Discussion}

Representative images and calculated EM distributions for the loops in our sample are shown in Figs. \ref{fig1} and \ref{fig4}. The electron density, peak temperature, peak EM, and the Gaussian width of the loop cross-field intensity profile are presented in Table \ref{table}.
In all cases the computed EM distribution is a strongly peaked function of temperature.
The narrow temperature distributions are consistent with previous results using spectroscopic data \citep{delzanna_2003,warren_etal2008a} and more limited imaging data \citep{aschwanden&nightingale_2005}, though the coverage of the imager observations was not sufficient to produce convincing temperature measurements. None of these studies were able to say anything definitive about possible substructure, but the very narrow temperature distributions measured for these loops suggest that they may consist of single monolithic structures. From Equation \ref{equation1} we see that we can use the EIS measurements of the density and temperature to infer the physical radius of the loop for the case of a single thread ($N$=1), which is $r_1$ in Table \ref{table}. We emphasize that measuring the density is critical for probing the internal structure of a loop since it allows us to isolate the geometrical factors that give rise to the observed emission.

\begin{figure}[t!]
\centerline{%
\includegraphics[viewport=20 240 570 650,clip,scale=0.5]{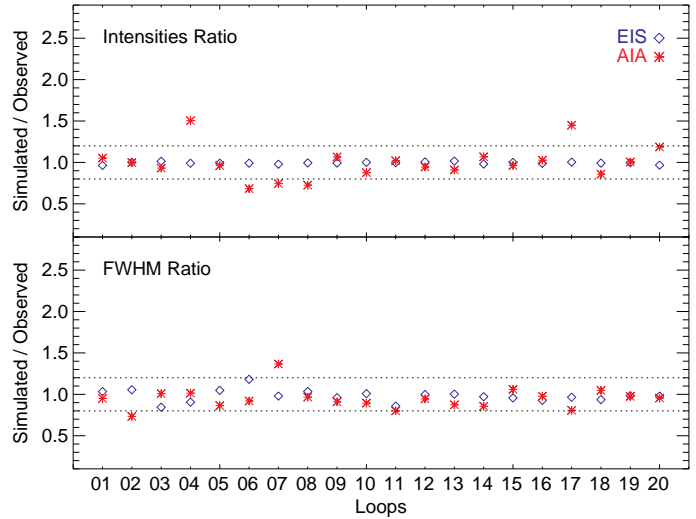}}
\caption{Ratios of simulated and observed intensities and FWHMs for both the 
EIS and AIA measurements for the entire loop sample. The simulations are able
to reproduce the observed values to within 20\% for most of the loops.
}
\label{fig3}
\end{figure}

The application of this procedure shows that for a minority of the cases that we have studied the EIS and AIA loop widths predicted from a single thread model are able to reproduce the observations. An example is loop \#19 which is shown in Figs. \ref{fig1}--\ref{fig2}. This loop is compatible with a simulation of a single thread with a radius of 830 km, and our emission measure analysis indicates that it is very nearly isothermal. Our results are the first to demonstrate unambiguously that some coronal loops can be resolved with current solar instrumentation. A previous study \citep{aschwanden&nightingale_2005} showed that the observed loop FWHMs are often larger than the instrumental PSF, as we also find here. This analysis, however, lacked density measurements and a model to deconvolve the physical size of the loops from the observed widths and instrumental PSF. As the authors of that paper pointed out, their analysis assumed a filling factor of unity, which is inconsistent with the measured values for coronal loops \citep{warren_etal2008a}. In a follow-up paper, \citet{aschwanden_etal2007a} did model the sub-resolution and considered broadening due to the PSF, but, without a density diagnostic, could not measure the volume of emitting plasma in coronal loops. 

For the majority of the loops in our sample, however, a single thread model produces simulated loops that are much narrower than what is observed. Such loops require substructure to reproduce the measured FWHMs. For these loops we have determined the minimum number of threads compatible with both the EIS and AIA data. The number of threads and the thread radius is given in Table \ref{table}. For most cases, the data are modeled to within the uncertainty of the instruments' calibration (20--25\%). The ratio of the simulated and observed intensity and loop FWHMs for each loop is summarized in Fig. \ref{fig3}. Since the multi-thread model has been constrained by the EIS observations the good agreement between the EIS measurements and the model is expected. The model parameters, however, are not derived from the AIA data and this comparison serves as an independent confirmation. As a further consistency check, we have verified that there is good agreement (within 10\%) between the measured FWHM for \ion{Fe}{12} 195.119\,\AA\, and the FWHM of the other lines. The models also still reproduce the intensities and FWHMs if the density is perturbed by the uncertainty in agreement between observed and calculated intensities ($\sim$ 35\%). Significantly, although the strand radius may change by this amount, the number of threads remains the same and the size of the loop envelope does not change appreciably.

\begin{figure}[t!]
\centerline{%
\includegraphics[viewport=150 130 440 750,clip,scale=0.85]{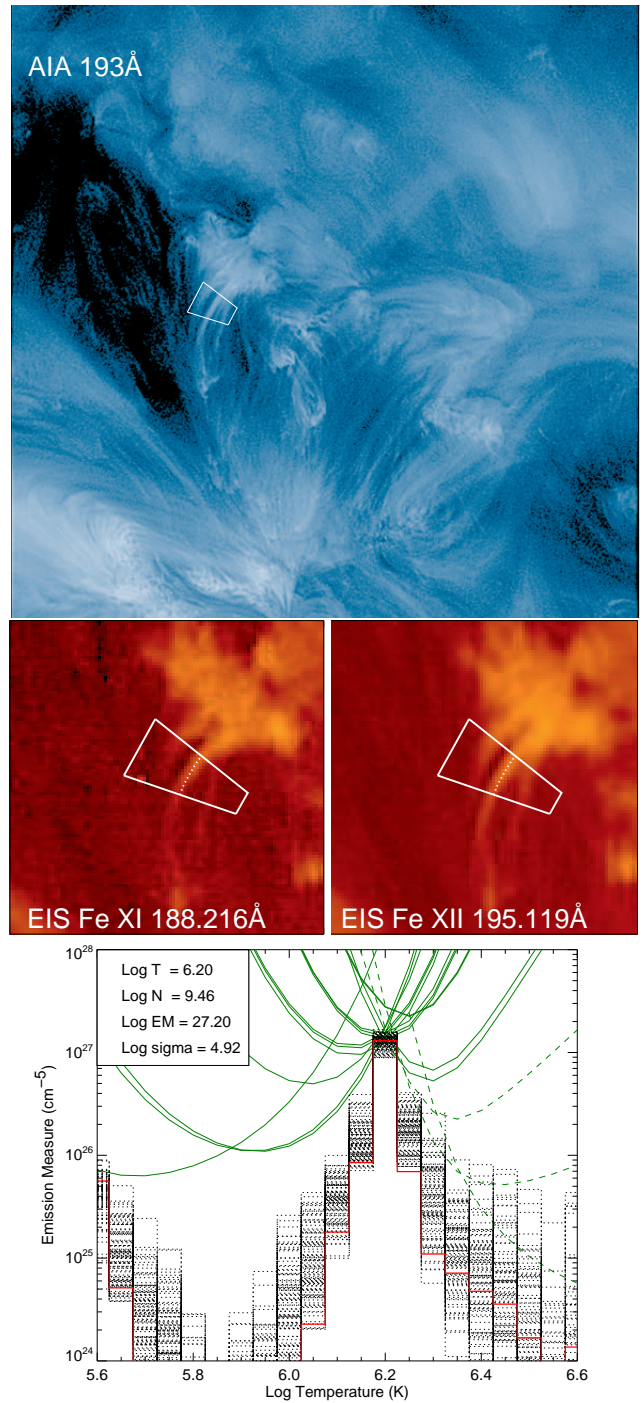}}
\caption{AIA 193\,\AA\, filter image (upper panel) and EIS slit raster scans in
\ion{Fe}{11} 188.216\,\AA and \ion{Fe}{12} 195.119\,\AA\, (lower panels) showing an 
unresolved loop. This is 
loop $\#$14 in Table \ref{table}. The traced
loop segments are indicated. The EM analysis of this loop is shown in the lower panel.
The description of this panel is the same as that of Figure \ref{fig1}.
}
\label{fig4}
\end{figure}
An unresolved loop (\#14) is shown in Figs. \ref{fig2} and \ref{fig4}. Here 5 threads with a maximum radius of 280 km can reproduce its observed properties and the emission measure analysis indicates that it has a finite but narrow thermal width, consistent with it being composed of a few threads. This loop is more typical of the majority of the sample. Our observations and models suggest, however, that the true size scale of even these coronal loops is not far from being resolved. A maximum of only 8 threads is needed to reproduce all of the loops in our sample, and the mean maximum thread radius of 480 km is not much smaller than the 720 km spatial pixels of EIS. Even the loops in our sample with the thinnest threads could be resolved with an instrumental spatial resolution of 200 km. This seems an achievable goal and we suggest that these results could be a reference to justify the spatial resolution required for a next generation spectrometer. Such an instrument would be able to routinely resolve coronal loops for the first time.

We stress that there is some ambiguity in the interpretation of these spatial scales since we do not know the maximum number of threads or the minimum thread radius. Loop \#14 in the figures is an intriguing example, however, since the EIS intensity profile is reproduced with relatively few threads, that are then seen to separate out when observed by AIA. It is difficult to justify large numbers of threads if the profile is not smooth. Furthermore, if they have to be constrained to specific positions the profile would be dependent on the viewing angle. Observations from the STEREO A and B spacecraft do not show any evidence that loop properties are line-of-sight dependent \citep{aschwanden_etal2008}.

Furthermore, our estimates of the loop width ($r$, $r_1$) are also supported by very high spatial resolution (100 km) optical observations of ``coronal rain'', which are condensations that form during the catastrophic cooling of coronal loops \citep{antolin&rouppevandervoort_2012}. Such events may be rare, but they offer unique insights into the structure of coronal loops. These observations yield a typical FWHM of 310 km, equivalent to a completely filled loop with a 215km radius when convolved with the instrumental PSF. They also show that neighboring cooling features occur nearly simultaneously, implying a coherence to the evolution of separate threads that is consistent with the narrow EM distributions we find. Such narrow distributions also suggest that the number of threads is small. An important avenue of future research is to investigate the evolution of the resolved loops to see if they are consistent with simple hydrodynamic models, or are heated quasi-steadily \citep{brooks&warren_2009,mulumoore_etal2011}.

Our observations present a challenge to current views on the fine magnetic coupling of loops to the photosphere, and heating models based on Parker's idea of nanoflare reconnection of braided magnetic fields in the corona \citep{parker_1983}. Typical magnetic flux and field strengths suggest $\sim$100 km radius photospheric flux tubes \citep{priest_etal2002}, and together with the very small spatial scales expected for magnetic reconnection, suggest that coronal loops observed at 1000 km scales should be composed of many hundreds of independent threads in various stages of heating and cooling \citep{klimchuk_2006}. This implies a broad distribution of temperatures at the spatial resolution achieved by EIS and AIA \citep{schmelz_etal2011}. Localized area expansion of the magnetic field could bridge the gap between photospheric and coronal spatial scales \citep{warren_etal2010a}, but there is a clear need to investigate the mapping of specific loops to the measured magnetic field. Finally, our observations of resolved single loops and EIS bundles reproduced by a few threads that split out into spatially separated individual structures when viewed by AIA, suggest that either magnetic reconnection occurs on much larger spatial scales than previously imagined, is energetically unimportant for loop heating, or the magnetic field in the corona is generally topologically simple and not braided.

\acknowledgements
We would like to thank the referee for their positive comments and suggestions.
This work was 
performed under contract with the Naval Research Laboratory and was funded by
the NASA {\it Hinode} program.
{\it Hinode} is a Japanese mission developed and launched by ISAS/JAXA,
with NAOJ as domestic partner and NASA and STFC (UK) as international partners.
it is operated by these agencies in co-operation with ESA and NSC (Norway).

\end{document}